\documentclass[12pt]{iopart}

\begin{document}

\title{Exploring non-signalling polytopes with negative probability}
\author{G. Oas$^{1,\dagger}$, J. Acacio de Barros$^{2}$, and C. Carvalhaes$^{3}$}

\address{$^{1}$ SPCS, 220 Panama Street, Stanford University, Stanford, CA
94305-4101}

\address{$^{2}$ LS Program, 1600 Holloway Street, San Francisco State University,
San Francisco, CA 94132}

\address{$^{3}$ CSLI, 220 Panama Street, Stanford University, Stanford, CA
94305-4115}

\ead{$^{\dagger}$oas@stanford.edu}

\begin{abstract}
Bipartite and tripartite EPR-Bell type systems are examined via joint quasi-probability
distributions where probabilities are permitted to be negative. 
It is shown that such distributions exist only when the no-signalling condition is satisfied.
A characteristic measure, the probability mass, is introduced and, via its minimization, limits
the number of quasi-distributions describing a given marginal probability distribution.
The minimized probability mass is shown to be an alternative way to characterize non-local systems.
Non-signalling polytopes for two to eight settings in the bipartite scenario
are examined and compared to prior work.
Examining perfect cloning of non-local systems within the tripartite scenario suggests 
defining two categories of signalling. 
It is seen that many properties of 
non-local systems can be efficiently described by quasi-probability theory.
\end{abstract}


\maketitle

\section{Introduction}

Ever since Bell's seminal work \cite{bell_einstein-podolsky-rosen_1964}, it has been clear that certain quantum mechanical sets of observables are stochastically incomplete, in the sense that no proper joint probability distribution exists that explain their expectations and moments. Later, Pitowski showed that classical systems, as defined by Bell, formed a polytope
within the space of probabilities, determined by CHSH-like inequalities \cite{pitowsky_quantum_1989}. Much work has been done ever since to understand why quantum mechanics violates classical probability, and, more importantly, why its boundaries are the way they are. For example, Popescu and Rohrlich asked whether non-signalling between two actors in a bipartite system could account for the observed quantum correlations \cite{popescu_quantum_1994}. They answered this question in the negative, by describing a range of states that do not signal and yet can not be described by quantum theory. 

To explore those boundaries, here we focus on bipartite and tripartite EPR-type systems. In a bipartite scenario, two observers, Alice and Bob, each receive one of two subsystems. Alice, has a choice of input $x \in \{0,..., n_x-1\}$ and a measurement yields outcome $a\in \{0,...,n_a-1\}$, while Bob inputs $y\in \{0,... ,n_y-1\}$ and receives output $b\in \{0,..., n_b-1\}$; such systems are labeled  $n_xn_yn_an_b$. This is viewed as Alice and Bob having a choice of random variables $\mathbf{A}_x, \mathbf{B}_y$ which can take one of the $n_a -1, n_b-1$ values respectively\footnote{Here we use the convention, standard in quantum computation, of having indices start with $0$ and not with $1$, i.e. $x \in \{ 0,\ldots ,N-1\}$, and not $\{1,\ldots ,N\}$.}. For example,  $2222$ is the label for standard Bohm-EPR system.

A question of interest is what type of correlations can be formed with various constraints imposed. Three regions can be clearly defined. First, from Pitowski's work we have the local polytope, $\mathcal{L}$. The second region is defined by the range of systems whose marginal probabilities can be described by states and operators on Hilbert space, the quantum set, $\mathcal{Q}$. Finally, we have a third region, $\mathcal{NS}$, satisfying the non-signalling condition (NS). NS states that marginal probabilities restricted to one observer do not change when 
 the context of the experiment is changed at a space-like separated event, i.e. 
\begin{eqnarray}
P(a|x_i) &=& \sum_b P(a,b|x_i,y_j) =  \sum_{b'} P(a,b'|x_i,y_k), \nonumber \\
P(b|y_i) &=& \sum_a P(a, b|x_j,y_i) =  \sum_{a'} P(a,'b|x_k,y_i), \forall i, j\neq k. \label{NS}
\end{eqnarray}
It is clear that $\mathcal{L}\subset\mathcal{Q}\subset\mathcal{NS}$, i.e. there are quantum systems that are non-local and there are non-signalling systems that are not quantum  \cite{popescu_quantum_1994}. 
 The commonly called `no-signalling' polytope is defined here as $\mathcal{P} = \mathcal{NS}\bigcap\mathcal{M}$, where $\mathcal{M}$ is the set of non-negative marginal probability
 distributions.

As an example, let us examine the $2222$ case, where non-locality is observed when Alice and Bob have binary inputs and outputs. It can be proved \cite{fine_hidden_1982} that a joint probability distribution over all random variables exists if NS and the following eight CHSH inequalities  \cite{clauser_proposed_1969} are satisfied:
\begin{eqnarray}
|S| &=& |\langle \mathbf{A}_0\mathbf{B}_0\rangle  + \langle \mathbf{A}_0\mathbf{B}_1\rangle  + \langle \mathbf{A}_1\mathbf{B}_0\rangle  - \langle \mathbf{A}_1\mathbf{B}_1\rangle | \leq 2, \label{CHSH}
\end{eqnarray}
with the other inequalities found by distributing the minus sign through the terms,
and  $\langle \mathbf{A}_i\mathbf{B}_j\rangle = P(\mathbf{A}_i = \mathbf{B}_j) - P(\mathbf{A}_i\neq \mathbf{B}_j)$.
Marginal distributions satisfying all eight inequalities, and satisfying NS, admit 
pre-established strategies or local hidden variable models. The local polytope is formed by the eight CHSH-saturating 7-dimensional facets, having 16 vertices. Each vertex corresponds to a joint probability distribution (jpd) where one  atom has probability 1. The quantum set extends beyond $\mathcal{L}$ up to a maximum described by  Tsirelson  $|S| = 2\sqrt 2$ \cite{cirelson_quantum_1980}. The non-signalling polytope contains the 16 local vertices and eight non-local vertices, where $|S|=4$. These non-local vertices are known as PR boxes \cite{popescu_quantum_1994}. 
The nature of the $2222$ system has been extensively studied and forms the benchmark to explore more complex scenarios.
Increasing the number of inputs, outputs, or parties leads to an explosion in number of Bell inequalities (facets) and vertices. 
It is a computationally hard problem to generate facets for systems with much more than 2 inputs. 

Usually unrelated to this discussion, many physicists proposed the use of modified probability theories to explain certain quantum phenomena. The quantum logic approach modifies the underlying algebra of sets \cite{pitowsky_quantum_1989}, while approaches such as general probabilistic theory modify the Boolean algebra \cite{barrett_information_2007}. Another strategy, considered here, is to maintain the Boolean algebra but modify the probability measure by allowing it to take negative values (see \cite{muckenheim_review_1986} for a comprehensive review). Such an approach is termed quasi-probability theory \cite{wigner_quantum_1932}, or simply negative probabilities \cite{dirac_bakerian_1942,feynman_negative_1987}. 

One of the often mentioned difficulties about negative probabilities is how to interpret them. For instance, both Dirac and Feynman avoided talking about meanings for negative probabilities, and proposed them only as a computational tool. However, recently this situation has changed. For instance, Andrei Khrennikov showed that if we take the von Mises frequentist interpretation of probabilities in terms of infinite sequences, negative probabilities are a consequence of the violation, in p-adic statistics, of the principle of statistical stabilization \cite{khrennikov1993p, khrennikov1993statistical, khrennikov1994p, khrennikov1994discrete, khrennikov1993statistical}. Also in the lines of a a frequentist interpretation, Samson Abramsky and Adam Brandenburger showed that positive and negative measures could be mapped into a negative joint probability distribution \cite{abramsky_operational_2014}. Finally, on the subjective side, a possible interpretation in terms of non-monotonic upper probabilities was proposed in \cite{de2014decision}. However, independent of interpretational issues, as Feynman mentioned in \cite{feynman_negative_1987}, no clear advantage seems to come from using negative probabilities. 

Despite Feynman's views, negative probabilities have seen a resurgence in the physics literature,  particularly in connection with super-quantal correlations \cite{han_y._explicit_1996,rothman_hidden_2001,halliwell_negative_2013,al-safi_simulating_2013,abramsky_operational_2014}. Also, inspired by the burgeoning field of quantum interactions, negative probabilities has been applied to context-dependent problems in the social sciences\footnote{The reader interested in more details about this field is directed to references \cite{khrennikov2010ubiquitous, busemeyer2012quantum, haven2013quantum}.}. For example, in decision making, violations of Savage's Sure-Thing Principle have been associated to negative probabilities \cite{khrennikov2010ubiquitous, haven2013quantum, de2012joint}. Because of the connection between negative probabilities and quantum interference \cite{deBarros2014negative}, a possible neural mechanisms have been proposed that could account for such violations \cite{suppes2007quantum, suppes2012phase, de2012quantum}. 

In this paper, we show an application of negative probabilities to explore local and non-signalling polytopes in the EPR-Bell scenario.
We start with a definition of negative probabilities that imposes further constraints than previous approaches, and show that such probabilities define non-signalling polytopes. We then examine the use of negative probabilities in bipartite systems of interest in quantum mechanics, and show a connection between negative probabilities and Bell-type inequalities. Finally, we investigate a tripartite system, and show that NS needs to be generalized in the case of quantum cloning.

\section{Quasi-probability}

As mentioned above, not all quantum systems allow for the existence of proper probability distributions. To overcome this, some authors proposed the relaxation of the non-negativity requirement for probabilities, thus leading to negative probabilities, or joint quasi-probability distributions (jqpd). However, one of the problems of relaxing this requirement is the explosion of possible jqpd's that are consistent with the observed expectations and moments. 

To limit the number of jqpds, two constraints are imposed.
The first is simply to reiterate the criterion imposed by Feynman in \cite{feynman_negative_1987}: any marginal probability derived from a jqpd
for which the outcome event can be realized within an experiment, what will be termed ``observable"
probabilities, must be non-negative. The second constraint is to minimize the 
{\em probability mass}, defined in this work as $M=\sum |p_i |$, where $p_i$'s are probabilities of atomic events \cite{de2014decision}\footnote{Our use of ``probability mass'' is not standard in probability theory, and whenever we refer to probability mass in this paper we mean it as given by the sum of absolute values of the (perhaps negative) probabilities of elementary events or atoms.}. This minimization is performed such that the observed marginal probabilities are preserved, restricting the jqpd to those with the lowest value for $M$ (L1 norm). Such minimized probability masses are labeled  $M^*$. With these constraints, states within $\mathcal{L}$ correspond to $M^*=1$, and $M^* > 1$ can be seen as a measure of deviation from a joint probability distribution.

Joint quasi-probability distributions are related, in a non-trivial way, to NS through the following theorem
(independently derived by \cite{al-safi_simulating_2013,abramsky_operational_2014}).
\begin{quote}
{\bf Theorem: } A necessary and sufficient condition for a system to satisfy the no-signalling condition (\ref{NS})
is the existence of a normalized joint quasi-probability distribution yielding the marginal probabilities of the system.
\end{quote}
{\em Sketch of the proof:} For sufficiency, if a joint exists, we 
express the NS condition (\ref{NS}) in terms of probabilities $p_{a_0a_1\ldots b_0 \ldots b_{d-1}}$ of the atoms $a_0a_1\ldots b_0 \ldots b_{d-1}$, and obtain
\begin{eqnarray}
P(a_i|x_iy_j) &=& \sum_{a_j\neq a_i} \sum_{b_j}  p_{a_0a_1\ldots a_i \ldots b_0 \ldots b_{d-1}} = P(a_i|x_iy_k), \label{4}
\end{eqnarray}
which is equivalent to NS. For necessity, we note that given a set of non-signalling observable marginals, the joint probability is related to them by a transformation $q=Ap$, where $q$ is a vector representing the marginals and $p$ the jqpd. Since $A$ is either full rank invertible (if we have all marginals) or rank deficient (depending on the number of observable marginals), a solution for $p$ always exists (though not necessarily non-negative or unique).

\section{Bipartite systems in terms of jqpds}

We return to the 2222 case and examine the range of jqpds. The CHSH inequalities \eref{CHSH} can be expressed in terms
of the probabilities of the 16 atoms, $p_{a_0a_1b_0b_1}$, as $S_{m,n} = 2\sum (-1)^{f_{n,m}}p_{a_0a_1b_0b_1}$
where $f_{m,n} = (a_0\oplus a_1)(b_0\oplus b_1) \oplus a_n\oplus b_m$, and $\oplus$ signifies addition modulo 2. 
For a particular inequality, with fixed $m$ and $n$, we label those probabilities corresponding to $f_{m,n}=0$
as $p$ and those corresponding to $f_{m,n}=1$ as $q$. Furthermore, we split
each probability into positive and negative parts, $p_i= p^+_i -p^-_i$, where $p_i^\pm,q_i^\pm, \geq 0$, and sum over each category,
$p^{\pm} = \sum p^{\pm}_{a_0a_1b_0b_1}, q^{\pm} = \sum q^{\pm}_{a_0a_1b_0b_1}$.
We then have $1 = p^+ - p^- +q^+-q^-$, $M = p^+ + p^- +q^++q^-$, and    
above each $S>2$ facet,
\begin{eqnarray}
M &=& \frac S2 + 2(p^- + q^+).   
\end{eqnarray}
The minimum value of the probability mass, $M^*$, obtains when $p^- = q^+ =0$.
This suggests the use of $M^* = |S|/2$ as a measure of departure from a local system.
As $M^*$ is easily computable for more complex scenarios, it is conjectured that it provides
 a general, more economical, approach to characterize local and non-signaling polytopes.

As an example, the well-studied isotropic systems, those having vanishing means and running from $S=0$ to a maximally nonlocal vertex (PR box),
can be parameterized as $p_i = \frac 1{16} + \frac x8, q_i =  \frac 1{16} - \frac x8, x\in (0,1).$ 
The box at $\partial\mathcal{L}$ occurs at $x = \frac 12$, $\partial\mathcal{Q}$ (at $S = 2\sqrt{2}$) at $x =  {1\over \sqrt 2}$,
and the PR box occurs at $x = 1$ where $M^*=2$.


\subsection{NN22 systems}

We now generalize to $N$ settings and examine  specific cases.
When moving from the $2222$ to $3322$ scenario, one new class of inequalities, $I_{3322}$, appears in addition to the CHSH inequalities. This class was generalized to $I_{NN22}$ \cite{collins_relevant_2004}. In the notation of \cite{brunner_bell-type_2006} it reads $\langle I_{NN22}|P\rangle \leq 0$ if $P\in\mathcal{L}$, where
\begin{eqnarray}
I_{NN22} &=& 
 \begin{array}{c|cccccc}  
       &  -1& 0 & 0 &\cdots & 0 & 0 \\ \hline
      -(N-1)& 1 & 1& 1 & \cdots & 1 & 1\\
      -(N-2)& 1& 1& 1 &\cdots & 1& -1\\
      \vdots & \vdots &  &  \\
      -1 & 1& 1& -1 &\cdots & 0& 0\\
	0 & 1& -1& 0 &\cdots & 0& 0\\
   \end{array}\label{INN22}
\end{eqnarray}
is the table of coefficients  whose entries are multiplied by the corresponding 
marginal probabilities in the following table
\begin{eqnarray}
P&=&    \begin{array}{c|ccc}  
       &  p(b_i = 0) \\ \hline
      p(a_j=0) & p(a_j=0,b_i=0)\\
   \end{array}  \;\; i,j = 1,\cdots ,N .
\end{eqnarray}

The inequality can also be expressed as a $2^{2N}\times 2^{2N}$ table of coefficients multiplying each probability of an atom.
For sake of brevity this table is not displayed; however, it is noted that a finite set of coefficients appear, $0, -1, -2, \cdots, -k$ where $k = \frac 12 N(N-1)$.
To relate $M$ to $I_{NN22}$ the probabilities of atoms are re-expressed, as above, according to their coefficient in the inequality, $p, q_1, q_2, \cdots, q_k$,
where $p$ are those corresponding to $0$ coefficients and $q_j$ to the $-j$ terms. As before, each probability of an atom is split into positive and negative terms
and each category is summed.

With this parameterization we have,
\begin{eqnarray}
1 &=& p^+ - p^- + \sum_{j = 1}^k (q^+_j - q^-_j), \label{Nnorm}\\
M &=& p^+ + p^- + \sum_{j = 1}^k (q^+_j + q^-_j),  \label{Nprobmass}\\
I_{NN22} &=&  \sum_{j = 1}^k (q^-_j - q^+_j).  \label{INN22inequal}
\end{eqnarray}
A maximal violation of $I_{NN22} $ and $M^*$ requires $q^+_j = 0$  and  $p^- = 0$. This, along with (\ref{Nnorm})-(\ref{INN22inequal}), give
\begin{eqnarray}
M^* &=& 2I_{NN22} + 1 - 2\sum_{j= 2}^k (j-1)q^-_j.\label{MINN22}
\end{eqnarray}
As discussed in \cite{brunner_bell-type_2006} a general form for $PR_N$ boxes can be given based on the structure of \eref{INN22}.
For those $PR_N$ boxes, $I_{NN22} = \frac 12(N-1)$ and, with (\ref{MINN22}), gives
a maximum value of $M^* = N$. However, this upper limit is only guaranteed for $N=2$, since for higher $N$ the sum in  (\ref{MINN22}) is nonzero.

 The minimization of $M$ is a nonlinear optimization problem and can be cumbersome to compute. 
 However, by splitting each probability of an atom into positive and negative parts it can be recast as a linear-programming problem:
 minimize \eref{Nprobmass} such that marginal probabilities are obtained and the distribution is normalized \cite{wright_numerical_1999}. 
 In the minimization process, there exists a subroutine to adjust cases where both $p^+$ and $p^-$ are non-zero probabilities for an atom, however it was never called in this work. 
 
 For $N=3$ all 1344 vertices of the 3322 non-signalling polytope were generated and all
vertices  had $M^* = 2$. This is in contrast with the conclusions of \cite{brunner_bell-type_2006} where the vertices
were placed into 4 categories and it was shown that the 192 $PR_3$ boxes require at least two PR boxes to be simulated. Suggesting that the $PR_3$ boxes are a stronger non-local resource than the $PR$ boxes. The contrast with the result
found here needs further examination.

For $N=4$ 
all $2^{16}$ non-deterministic extremal boxes, those with $p(a_i) = p(b_j) = \frac 12, p(a_i,b_j) = 0$ or $\frac 12\; \forall i,j, $  fall
within 4 classes: 128 are local boxes ($M^* = 1$), 43904 with $M^* = 2$, 12288 with $M^* = 2.33$,
and 9216 boxes have the value $M^* = 2.4$.  Systems up to $N=8$ were generated and more classes of maximal vertices are
found with the maximum value increasing with additional settings, up to $M^*=2.9091$ for $N=8$.
Of significance is that  for $N>3$ vertices with $M^* = 2$ satisfy the $M_{NN22}$ inequality of \cite{brunner_bell-type_2006}
and can be simulated with a single $PR_{N-1}$ box, those vertices with $M^*>2$ are all $PR_N$ boxes. This suggests further classes 
amongst $PR_N$ boxes beyond that found in \cite{brunner_bell-type_2006}.

\section{Tripartite systems in terms of jqpds}

Here the most basic
tripartite systems is briefly examined in relation to perfectly cloned systems.
To implement perfect cloning, one needs to arrive at a joint conditional distribution that yields 
identical marginals, i.e., 
$p(abb'|xyy')  \longrightarrow  p(ab|xy) =  p(ab'|xy')$,
for all settings and outcomes.
It is known that all perfectly-cloned non-local systems permit signalling: if the original system is a PR box, which satisfies $a\oplus b = xy$, the cloned subsystem satisfies $a\oplus b' = xy'$, then
Bob can determine Alice's setting, $x$, since $a\oplus b \oplus a \oplus b' = xy \oplus xy'$, 
which yields $b\oplus b' = x(y\oplus y')\rightarrow (b\oplus b')(y\oplus y') = x$.

Recalling the parameterization of probability of atoms for isotropic boxes in the 2222 case,
a perfectly
cloned isotropic box can be represented by the following jqpd 
$p_{a_0a_1b_0b_1b'_0b'_1} = (\frac {6x+1}{64})(1-f_{0,0}) + (\frac {1-2x}{64})f_{0,0}$, where 
$f_{0,0} = (a_0\oplus a_1)(b_0b'_0 \oplus b_1b'_1)\oplus a_0\oplus b_0b'_0 
\oplus (a_0\oplus 1)(a_1\oplus 1)(b_0\oplus b'_0)\oplus a_1(a_0\oplus 1)(b_1\oplus b'_1)$.
Once outside of the local polytope, at least one observable marginal must be negative.  For example, the marginal
$P(a_0 = 0, b_0=1, b'_1 = 1)  = \frac 18(1-2x)$ and thus no non-local state can be perfectly cloned. Again, all of these
systems satisfy the no-signalling condition, but are known to allow signalling.

\section{Conclusion}

We showed that negative probabilities provided an additional tool to examine a variety of correlated bipartite and tripartite systems. For instance, a theorem relating the existence of jqpds to satisfaction of the NS condition was introduced. Furthermore, negative probabilities provide a more efficient method, compared to the ones existent in the literature, to examine non-signalling polytopes of any dimension, since they can be obtained from computationally fast linear-programming techniques. 
The method of extending standard probability theory introduced here brings a new view of non-local systems that has the
potential to simplify analyses of complex correlation polytopes and shed light on the nature of and relationship between entanglement,
non-locality, and contextuality.

For the  $2222$ system it was shown that  for non-local systems $M^* = \frac {|S|}2$,
allowing the former to be interpreted as a measure of deviation from a local system. 
This measure was then applied to cases up to eight settings for Alice and Bob. 
In the 3322 case all non-local vertices are found to have $M^* = 2$, 
while for 4422 three values are obtained, $M^* = 2, 2.33, 2.4$ with higher classes with increased settings. 
For $N>3$ vertices with $M^*=2$ fall under the $PR_{N-1}$ (or less) category while those with $M^*>2$ are
$PR_N$ boxes. This suggests previously uncharacterized differentiations amongst $PR_N$ boxes.

We also showed that for the perfectly cloned PR box, known to permit signalling, a jqpd exists, and, from our theorem,
NS is satisfied. 
This suggests two distinct forms of signalling for two space-like separated systems: an explicit one, where actions for one observer
change the marginal probabilities for the other (violating NS), and
a more passive form where information is revealed but not changed.
For the latter, Feynman's criterion is violated for perfect cloning of any non-local state.

It still remains to relate the minimal probability mass to concepts relating to non-local systems in a more precise way.
The view moving forward is that $M^*$ is more precisely a measure
of the contextuality embodied within a system. This requires a more careful study of jqpds and systems with contextuality, e.g. GHZ, Kochen-Specker type
scenarios, and recently discussed KCBS relations. To make progress, the approach here needs to be related to  an independent approach that explicitly embodies the contextual nature of the system, one such method is the 
contextuality by default approach \cite{dzhafarov_contextuality_2014,deBarros2014unifying}.

\paragraph{Acknowledgements.}
The authors wish to thank Professors Patrick Suppes, Charles de Leone, Ehtibar Dzhafarov, Janne Kujala, Stephan Hartmann, and Andrei Khrennikov for useful discussions about negative probabilities and quantum mechanics, as well as the anonymous referees for comments and suggestions.

\section*{References}

\bibliography{QuantumNegProb2}
\bibliographystyle{unsrt}

\end{document}